\documentclass[aps,pre,twocolumn,groupedaddress,showpacs,floatfix]{revtex4}
\usepackage{amsmath}
\usepackage{amssymb}
\usepackage{graphicx}

\begin{document}

\title{Visual Saliency and Attention as Random Walks on Complex Networks}

\author{Luciano da Fontoura Costa} 
\affiliation{Instituto de F\'{\i}sica de S\~ao Carlos. 
Universidade de S\~ ao Paulo, S\~{a}o Carlos, SP, PO Box 369,
13560-970, phone +55 16 3373 9858,FAX +55 16 3371 3616, Brazil,
luciano@if.sc.usp.br}

\date{2nd Feb 2007}

\begin{abstract}   
The current article shows how concepts from the areas of random walks,
Markov chains, complex networks and image analysis can be naturally
combined in order to provide a unified and biologically plausible
model relating saliency and visual attention.  Two types of models are
proposed: (i) images are converted into complex networks by
considering pixels as nodes while connections are established in terms
of fields of influence defined by visual features such as tangent
fields induced by gray-level contrasts and distance; and (ii) image
pixels exhibiting particularly distinctive values of visual properties
such as gray-level intensity, contrast, size of objects, orientation
and texture are mapped into nodes and the weights of links are defined
in order to favor transitions between regions with similar or
different visual features, also taking the distance between the nodes
into account.  Preferential random walks are performed on such
networks in order to emulate attentional shifts and eye movements, and
the saliency of each region is obtained in terms of the frequency of
visits to each node at equilibrium.  In the case of the first model,
there is a definite tendency to emphasize not only high curvature
points but also convergences of the tangent field.  The frequency of
visits is found to be strongly correlated with the node degrees
(strengths) for this model.  Different results have been obtained for
the second model as a consequence of the directed and asymmetric
nature of the respectively obtained networks.  
\end{abstract}


\maketitle

\emph{The ability to focus attention on important things is a 
defining characteristic of intelligence.}
(R. J. Shiller)
\vspace{0.2cm}

\section{Introduction}

Vision~\cite{Marr:1982} is the ability, given a specific scene, to
recognize the existing objects and their respective properties
(e.g. position, rotation, size, etc).  Although vision is natural to
animals, achieving maximum flexibility in primates, all attempts by
science and technology to emulate this ability have to a large extent
failed --- full fledged vision is simply too complex.  Artifacts
(e.g. shadows and occlusion), 2D projections, and noise always present
in images imply a degenerated mapping from real scenes to the
biological visual representation, so that the effective recognition of
objects ultimately demands high levels of intelligence and
comprehensive models of the visual features of our world.  Actually,
even the natural solutions to vision have been achieved at great cost
and difficulty.  Though nearly 50\% of the human cortex is dedicated
at varying degrees to visual analysis and integration, only a very
small region of the visual space, the area falling onto the
\emph{fovea}, can be  carefully analyzed at higher resolution by such a
formidable parallel computing system at any time.  Even so, the
remaining several limitations of vision are attested by a myriad of
optical illusions.

The serious limitations of the cortical hardware in processing vision
ultimately implied the retina to perform effective pre-processing in
order to filter out redundancies (luminance correlations) before
forwarding the visual information to the brain, via the lateral
geniculate nucleus~\cite{Tovee:1996}.  This is to a great extent
achieved through detection of the borders of the objects in images,
which tend to be associated to luminance contrasts.  Because only the
fovea, an area of the retina accountable for about just one degree of
the visual field, is engaged in high resolution image analysis, it is
important to have effective means for moving this small window along
time and space, performed through saccadic eye
movements~\cite{Tovee:1996}, so as to integrate along time and space
the most important portions (saliencies) of the image into a sensible
whole.  Extensive experimental investigations have shown that points
exhibiting high contrast
(e.g.~\cite{Parkhurst_etal:2002,Krieger_etal:2000}) and/or curvature
(e.g.~\cite{Attneave:1954}) tend to play a decisive role in saliency
definition and detection.  Other important experimental evidences
include the presence in the primary visual cortex of many neurons
which are orientation sensitive, exhibiting the so-called simple and
complex receptive fields, in the sense of being capable of estimating
the tangent field along the retinotopic representation of the
scene~\cite{Tovee:1996}.  Because of the decreasing resolution along
the retina as one moves from its center to the periphery, it is
reasonable to assume the saliency of local portions of the image to be
inversely related to the distance from those portions to the center of
the fovea (or attention).  In addition to gaze shift driven by
saliencies, more subtle visual mechanisms are performed on the
peripheral visual field in order to decide where to look next.  The
shifts of attention and their relation to saliencie saliencies
involving or not eye movements are the main subject of this article.

In spite of the intense and extensive experimental and theoretical
research in visual perception, relatively few physics-based approaches
have been proposed relating saliency detection and selective
attention.  In addition to the now classical work of Brockmann and
Geisel~\cite{Brockmann:2000}, who modeled human scanpaths in terms of
stochastic jumps in random quenched salience fields, a more recent
model of gaze shift has been reported by Boccignone and
Ferraro~\cite{Ferraro:2004}, who used non-local transition
probabilities over a previously estimated saliency field upon which a
constrained random walk was performed.  This work considered
transitions between the pre-assigned saliencies at any uniformly
random chosen orientation around each node.  The present work combines
recent results from the areas of complex networks
(e.g.~\cite{Albert_Barab:2002,Newman:2003,Boccaletti_etal:2006}),
Markov models (e.g.~\cite{Bremaud:2001}), random walks
(e.g.~\cite{Tadic:2001,Costa_Sporns:2006}), and artificial image
analysis (e.g.~\cite{Costa:2001,Costa_vision:2004}) in order to
develop a simple and yet unified and biologically plausible model
naturally integrating gaze shift and salience detection.  One of the
main theses supported in the current work is that

\begin{center}
  \fbox{$saliency \rightleftharpoons selective \ attention$}
\end{center}

i.e., saliencies and selective attention would be inherently
intertwined in the mammals visual system (and possibly also in
artificial vision).  The two types of models suggested in the current
article involve representing the image as a complex network such that
each part of the image (or object) is mapped as a node and connections
between these nodes are established in terms of specific visual
features such as the tangent field induced by the image contrasts,
distance between points/objects and size of objects, as well as
differences between the visual properties around each node.  The image
under analysis is henceforth represented in terms of the matrix $A$,
whose elements $(i,j)$ ($i = 1, 2, \ldots, N_x$ and $j= 1, 2, \ldots,
N_y$) are called \emph{pixels} (picture elements) and the gray-level
value $A(i,j)$ is proportional to the image luminance (only black and
white and gray-level images are considered here).

First, we address the case of selective attention driven by tangent
fields defined by the image gray-level variations.  Images can have
their borders detected (a procedure similar to that performed by the
retina) and a random walk performed along the herein defined tangent
field.  The steady state of the visits to nodes is conveniently
calculated from the eigenequation involving the stochastic
(transition) matrix associated to the respectively driven Markov
chain.  Interestingly, the saliencies of the image are naturally
revealed as a consequence of the most frequently visited nodes, at the
same time as these points act as beacons for the random walk,
therefore naturally integrating selective attention/gaze shift and
saliency manifestation.  The effects of having the connections in the
complex network representation of the image to be unit or inversely
proportional to the distance between pairs of points, as well as
random walks characterized by uniformly random choice of next moves or
preferential to the degree of the target nodes, are considered and
discussed.  Another interesting finding is the strong correlation
identified between the frequency of visits to nodes and the respective
degrees/strengths.  A second type of model for selective
attention/saliency is also presented in this work where the nodes are
pre-defined in terms of some specific visual property related to the
contrast, curvature, size and orientation of objects, as well as the
distance between the latter.  The transition probabilities between
nodes are defined in order to promote movements towards nodes
exhibiting similar or different visual attributes while being
inversely proportional to the distance between nodes (all considered
networks are geographical).

\section{First Model: Orientation Field Networks}

The first step in this type of model involves transforming the image
into a complex network $\Gamma$.  As in the retina, special attention
is given to the pixels defining high gray-level contrasts typically
found at the borders of objects.  These pixels can be conveniently
detected by using the Sobel or Laplacian of Gaussian filters
(e.g.~\cite{Marr:1982,Costa:2001}) for gradient estimation.  The
orientations defined by such edges (parallel to them) can be estimated
directly from the Sobel operator or by the spectral method for tangent
and curvature field estimation~\cite{Costa:2001}.  As the current
article is restricted to binary images, the latter method for tangent
field estimation has been applied.  The orientation of the edge at
position $(i,j)$ is henceforth represented as $\alpha(i,j)$.

The connections of the complex network $\Gamma$ representing the image
are established as follows: each detected edge element $(i,j)$ is
representedas a node and connected through unit and symmetric weights
with all other edge elements $(p,q)$ lying on the straight line
passing by $(i,j)$ with inclination $\alpha(i,j)$.  An alternative
connectivity scheme also considered in this work involves assigning
weights which are symmetric but instead of being unitary, are set as
being inversely proportional to the distance between the pixels
$(i,j)$ and $(p,q)$.  Note that the network connections therefore
correspond to visual associations induced by strong contrasts in the
image, accounting for the tendency of human vision to follow straight
lines.  Actually, this representation of an image in terms of a
complex network can be understood as a particular case of the
methodology suggested in~\cite{Costa_vision:2004}, where pixels are
connected while taking into account the similarity between their
properties (in the case of the current work, their tangent field).

\begin{figure}[h]
 \begin{center} 
   \includegraphics[scale=0.45,angle=0]{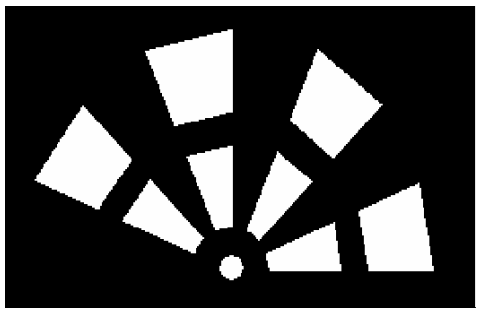} \\
   (a) \\ 
   \vspace{0.5cm}
   \includegraphics[scale=0.55,angle=0]{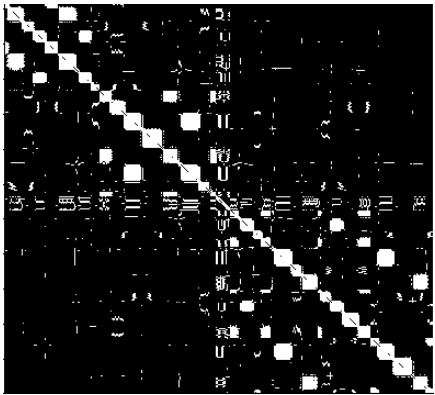} \\
   (b) \\
   \vspace{0.5cm} 
   \caption{The original binary image (a) and its respective weight
   matrix (c) obtained by `traditional' random walk with unit
   weights.~\label{fig:img}}
\end{center}
\end{figure}

Figure~\ref{fig:img} illustrates an image (a) as well as the weight
matrix (b) of the respectively associated complex network considering
unit weights.  The weight matrix considering strengths inversely
proportional to the distance between pairs of edge elements would be
visualized as a nearly diagonal matrix, so it is not shown.

Observe that, at this point, we have neither salience detection nor
attention dynamics implemented yet in our model yet.  In order to try
to obtain both these important concepts in a simple and integrated
way, traditional~\footnote{Namely random walks in which the next edge
to be trailed is chosen with uniform probability among the edges
connected to the current node.} and preferential random walks are
performed on the complex network.  The stochastic matrix $S_1$
associated to traditional random walks can be inferred from the
respective weight matrix $W$ of the complex network as

\begin{eqnarray}
  od(i)=\sum_k{W(k,i)}  \\
  S_1(i,j)=W(i,j)/od(j)
\end{eqnarray}

where $od(j)$ stands for the outdegree (our outstrength) of node $j$.
Preferential walks are alternatively performed so as to randomly
choose the next edge (among those emanating from the current node)
with probability directly proportional to the degree (or strength) of
the respective target nodes.  In this case, the respective stochastic
matrix becomes:

\begin{eqnarray}
  v(i) = \sum_k{od(k)} \ | \ W(k,i) \neq 0  \\
  S_w(i,j)=W(i,j)/v(j)
\end{eqnarray}

The outdegree (or outstrength, a term often applicable to weighted
networks) of a node $i$ therefore corresponds to the sum of all
weights of outbound edges along the column $i$ of the stochastic
matrix.  Both the above matrices necessarily follow the particular
eigenvalue equation $\vec{f}=S\vec{f}$, so that the frequency of
visits to nodes in the steady state is immediately given by the
respective values of the eigenvector $\vec{f}$ associated to the unit
eigenvalue, normalized so that $\sum_i f_i=1$.

\begin{figure*}[h]
 \begin{center} 
   \includegraphics[scale=0.45,angle=0]{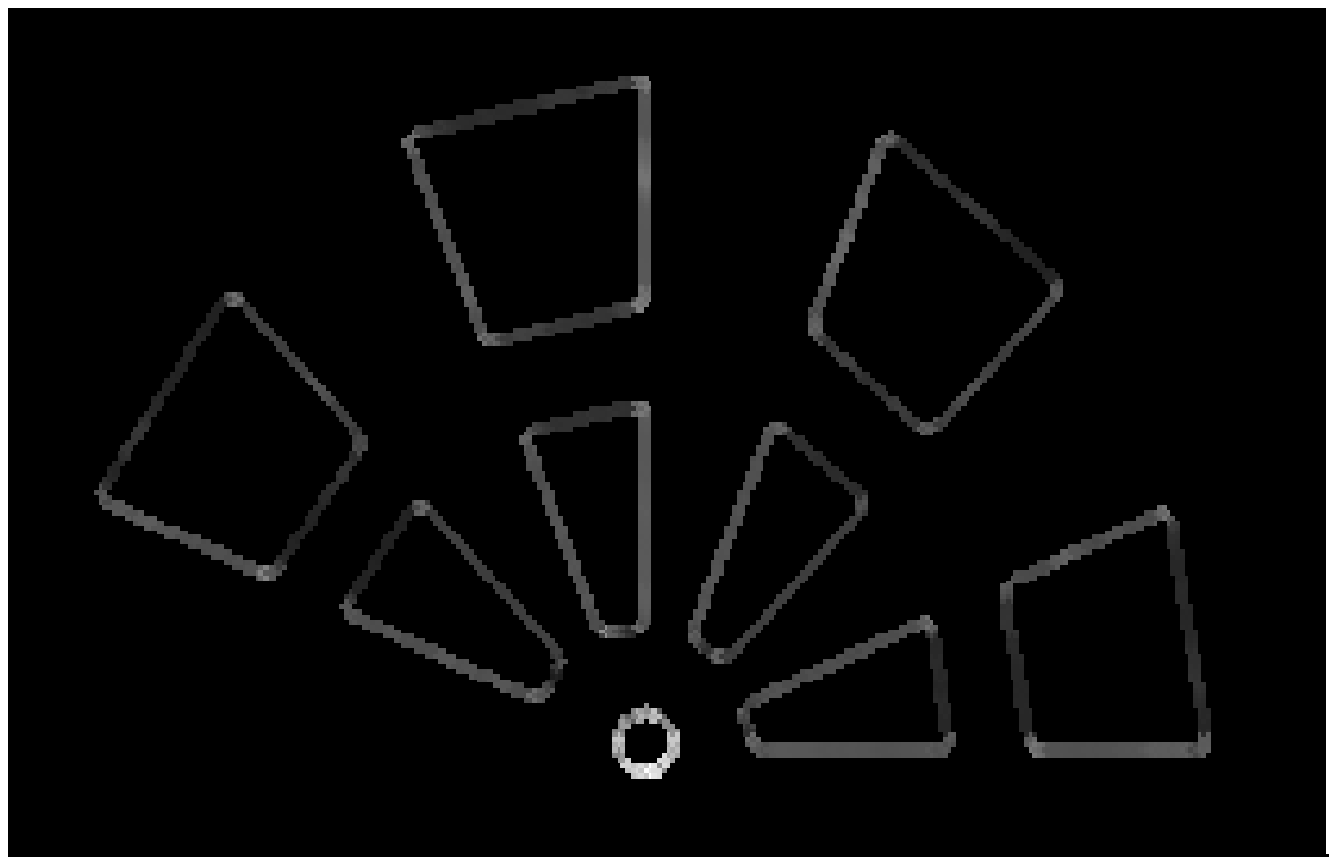} \hspace{1cm}
   \includegraphics[scale=0.45,angle=0]{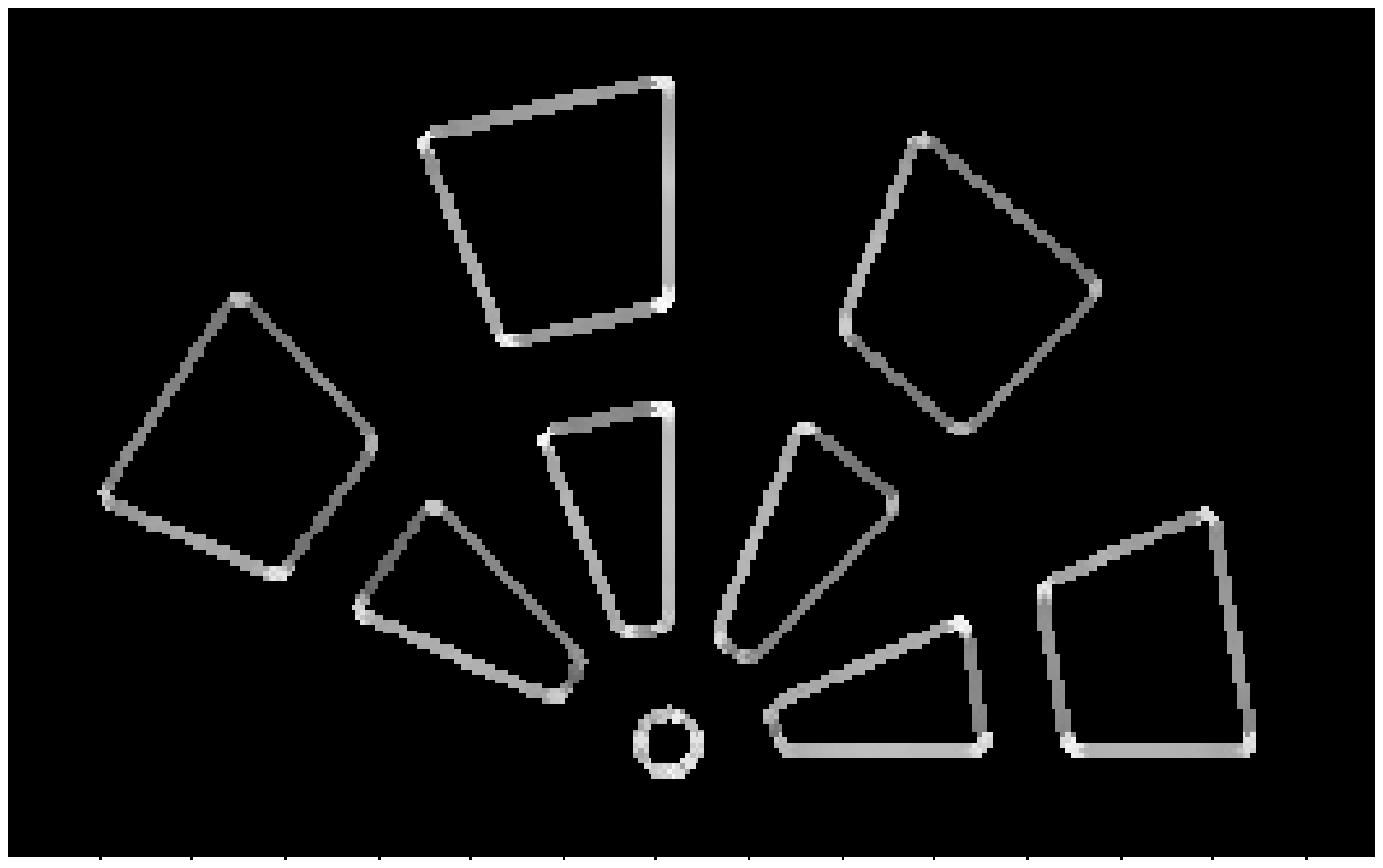} \\
   (a) \hspace{7.5cm} (b)\\
   \vspace{0.5cm}
   \includegraphics[scale=0.45,angle=0]{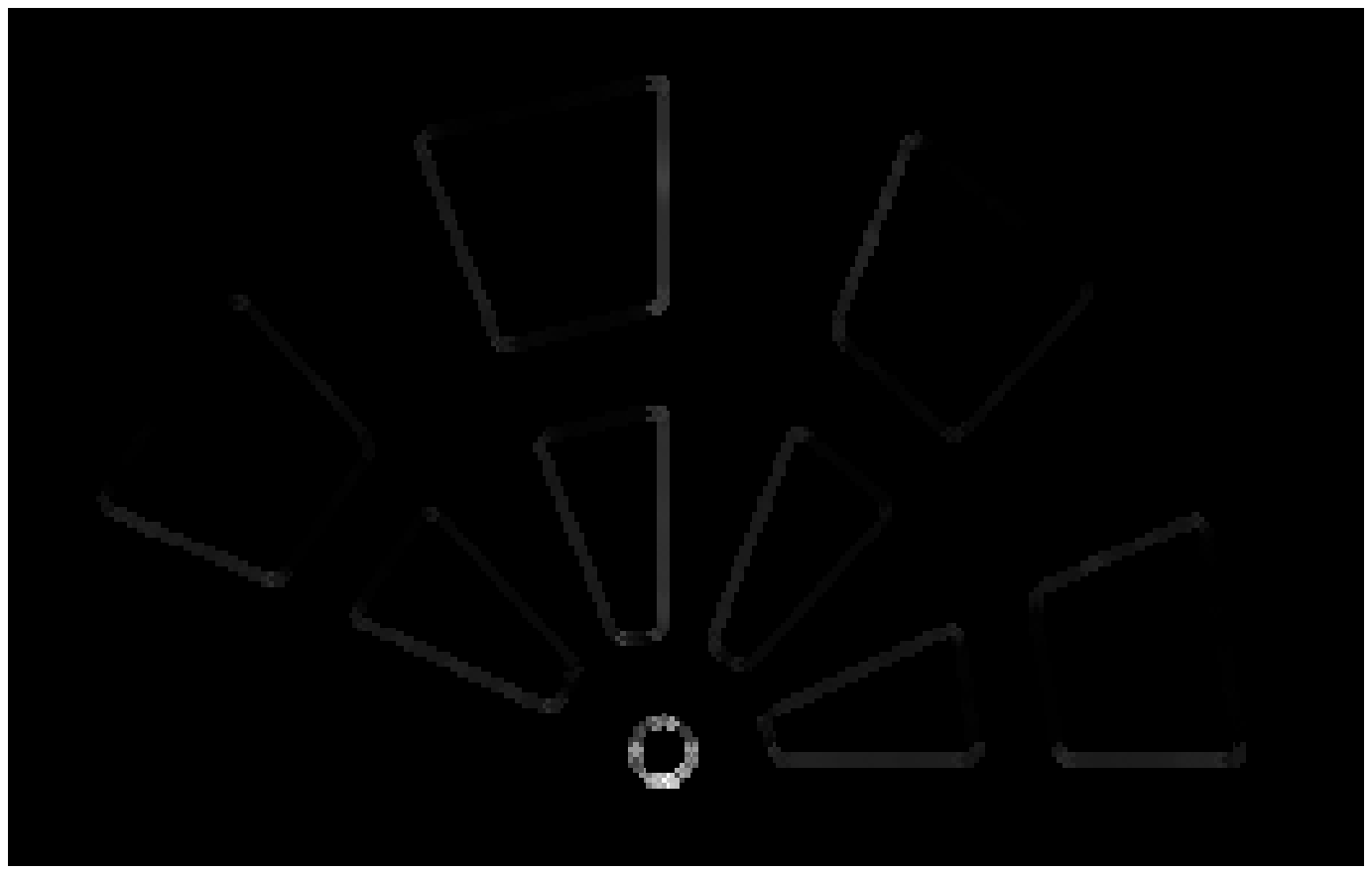} \hspace{1cm}
   \includegraphics[scale=0.45,angle=0]{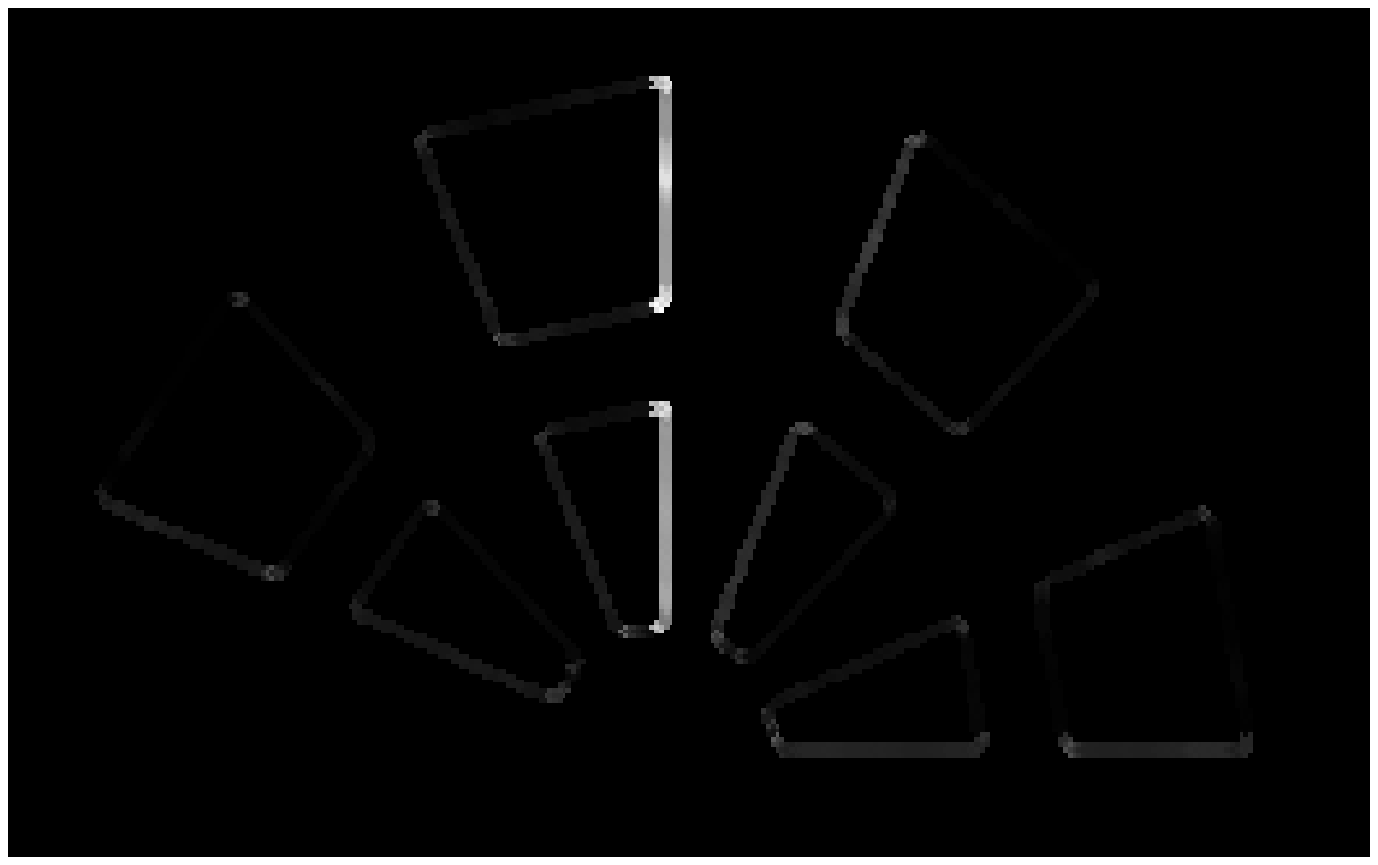} \\
   (c) \hspace{7.5cm} (d)\\
   \vspace{0.5cm}
   \caption{Saliency fields obtained for `traditional' random walks
   with unit (a) and inversely proportional (b) weights.  Preferential
   random walks considering the original image with (c) and without
   (d) the circular shape.~\label{fig:sali}}
\end{center}
\end{figure*}

While the random walks provide the means to mimic the tangent field
driven attention shift, the saliencies can be defined as the most
frequently visited nodes during a long walk, i.e. as a consequence of
the random walk dynamics. Figure~\ref{fig:sali} shows the obtained
saliency field for the image in Figure~\ref{fig:img}(a), assuming
`traditional' random walks. Higher activities are denoted by clearer
gray levels.  Interestingly, the saliencies not only tended to
correspond to high curvature points (vertices along the outlines of
the objects) but also resulted particularly marked at the convergence
of the tangents (i.e. into the circular shape at the bottom of the
figure).  The dominance of such a point of convergence ultimately
implied the remainder of the image to become darker, because of the
normalization of the normalization of gray-level variation adopted for
the sake of better visualization.  Also interesting is the fact that
longer lines, or sets of aligned lines, tend to be characterized by
higher frequency of visits to all involved nodes, suggesting another
dimension of the salience in visual structures, namely as being
proportional to the lenght of the aligned structures.  Another
important point to be born in mind is that the anysotropies of the
representation of objects into a orthogonal
(e.g.~\cite{Rosenfeld:2004,Costa:2001}) lattice imply the lines to
result with a degree of jaggedness which is dependent of the
respective line orientation, with the highest quality lines verified
for the horizontal and vertical orientations.  The frequency of visits
obtained by considering weights inversely proportional to the distance
between connected nodes are given in Figures~\ref{fig:sali}(b).
Although the vertices have again resulted as being particularly
salient, the convergence at the circular shape is much weaker than for
the `traditional walks', as a consequence of the fading effect of the
edges along distance.

\begin{figure}[h]
 \begin{center} 
   \includegraphics[scale=0.45,angle=0]{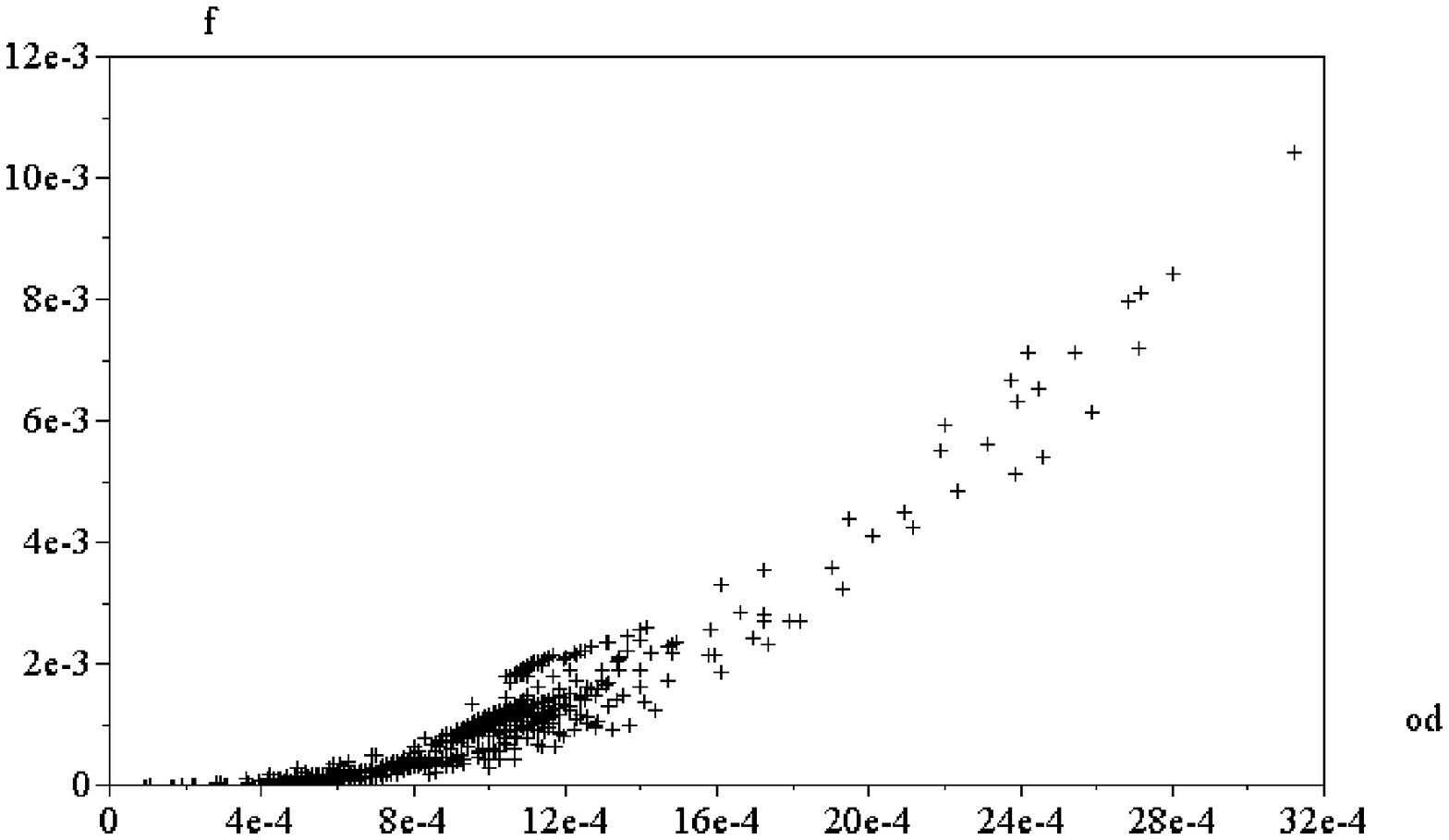} \\
   \vspace{0.3cm} 
   \caption{The superlinear correlation between the frequency of
   visits to pixels and their respective outdegree obtained for
   preferential random walks.~\label{fig:correl}}
\end{center}
\end{figure}

The results obtained for inversely proportional weights and random
walks with preferential choice of movements (favoring destination
nodes which have higher outstrength) are shown in
Figure~\ref{fig:sali}(c).  The interesting effect of this alternative
dynamics implied some elements of the image to become significantly
more salient, as is the case with the circular shape.
Figure~\ref{fig:sali}(d) shows the saliency field obtained while
disconsidering the circular shape.  The most salient figure now
corresponds to the vertical line.  It is particularly remarkable that
this line, and not the horizontal line (both of which presenting null
jaggedness), have been highlighted.  This is explained because the
vertical line is the subject of more crossings with other lines than
the horizontal one, therefore attracting more converging movements
emanating from the other parts of the image.  Figure~\ref{fig:correl}
shows the correlation between the frequency of visits $f$ and the
outdegree $od$ observed for the case shown in
Figure~\ref{fig:sali}(d).  It is clear from this result that pixels
which are topological hubs tend to be activity hubs also, in a
superlinear way, accounting for the enhancement of saliency implied by
the preferential random walks.  A strong linear correlation has been
observed for the traditional random walks, suggesting the use of
pixels outdegree as predictors of dynamical activity and saliency.

\section{Second Model: The Effect of Visual Properties on Saliency}

The second modeling approach considered in this work involves a
pre-selection of image regions and their mapping into respective
nodes, instead of considering each pixel as a node as in the previous
methodology.  The regions can be selected according to several
criteria.  For instance, regions of particularly high gray-level
values, contrast or curvature can be detected and mapped into nodes.
It is also possible to define a regular sampling of the image by a
lower resolution grid and consider specific features around each node,
such as gray-level intensity, contrast, orientation, spatial frequency
and even object size.  Once such regions are chosen, it is necessary
to define some means of obtaining the connectivity between them.  In
this work we consider the similarity between the visual features of
each pairs of nodes as well as the distance between those nodes.

More specifically, we define the weights $w(i,j)$ of the respectively
associated weight matrix as $w(i,j) = f(s(i),s(j))/d(i,j)$, where
$s(m)$ is the individual visual feature (or feature vector) associated
to each node $m$ and $d(i,j)$ expresses the distance between the nodes
$i$ and $j$; $f(s(i),s(j))$ is a function of the visual properties of
nodes $i$ and $j$, such as the Euclidean distance between those
measurements.  The transition matrix underlying the preferential
random walks is directly obtained from the weight matrix by a
normalization procedure implying all the colums of the transition
matrix to add to 1.

\begin{figure}[h]
 \begin{center} 
   \includegraphics[scale=0.35,angle=0]{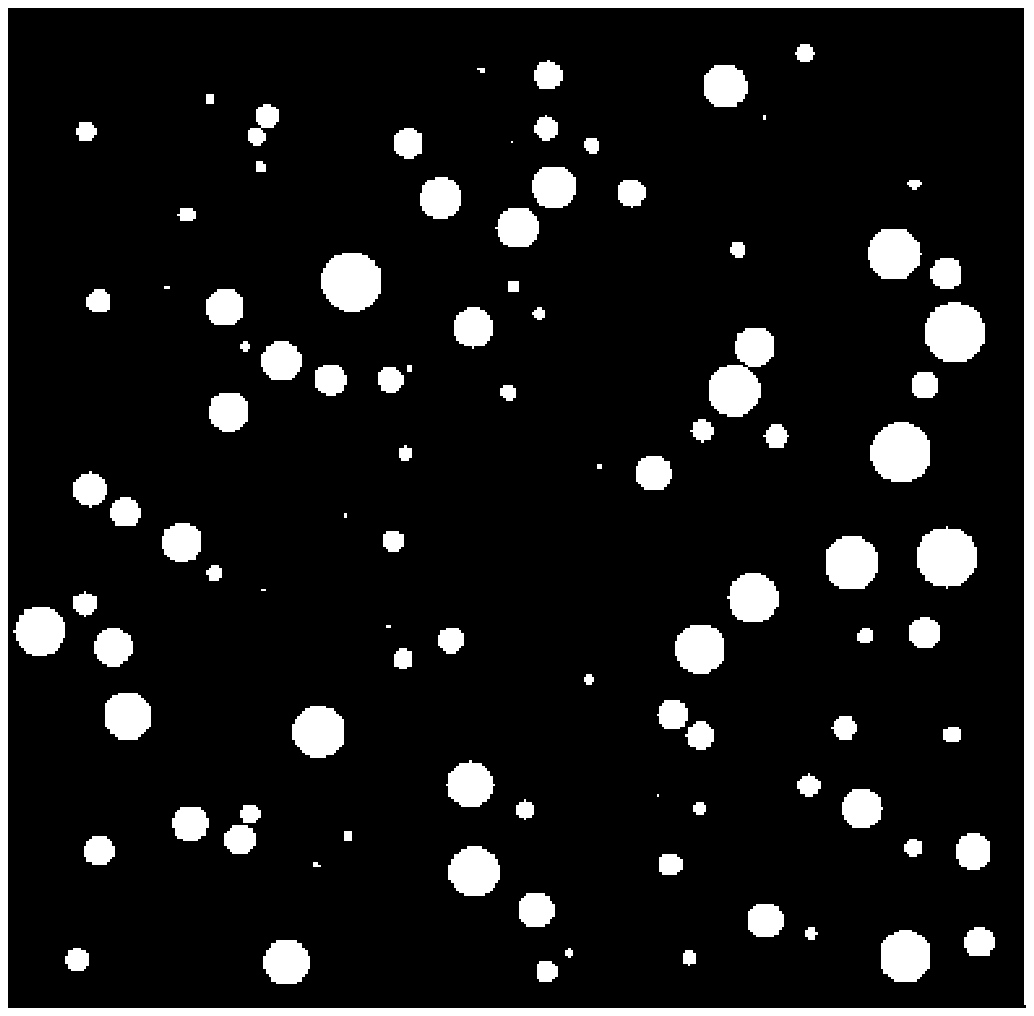} \hspace{1cm}
   \includegraphics[scale=0.35,angle=0]{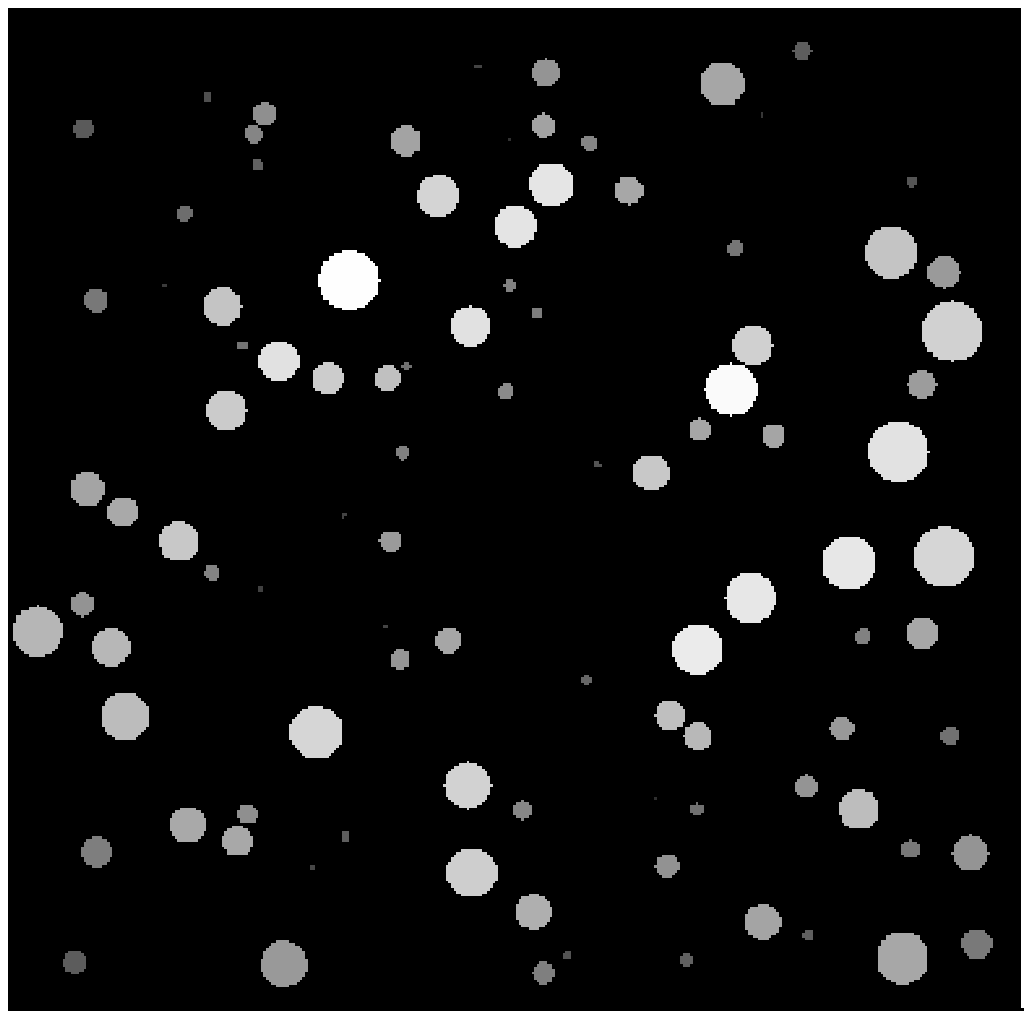} \\
   (a) \hspace{4.5cm}  (b) \\
   \vspace{0.3cm} 
   \caption{An $1000 \times 1000$ pixels image containing 100 disks
   with varying sizes (a), and the respectively obtained saliencies
   (b) expressed by the gray-level of the disks.~\label{fig:attn}}
\end{center}
\end{figure}

Figure~\ref{fig:attn} shows an image containing several disks (a) as
well as the respective saliency field (b) obtained by considering each
disk as a node and taking $f(s(i),s(j))=s(j)$, where $s(j)$ is the
radius of the destination node.  The resulting saliencies,
corresponding to the highly visited nodes during preferential random
walks, can be found to be directly related to the radius of the disks.
At the same time, nodes which are surrounded by a greater number of
nearby disks tend to become more salient, implying the objects at the
borders to be less salient.  We have found verying degrees of
correlation between the resulting saliency and the in- and
out-strength.  Similarly varying correlations have also been observed
between the resulting saliencies and the original visual features (the
radius disk, in the case of this example). 

\section{Concluding Remarks}

All in all, this work has proposed two models of selective attention
and salience detection founded on several concepts from the areas of
complex networks, Markov models, random walks, and artificial image
analysis.  These models are conceptually appealing because of the
natural integration of attention and salience definition, the
flexibility to be used at varying spatial scales and considering
different driven mechanisms, and the interesting correlations between
simulated activity and network topology.  The potential of the
approach has been illustrated with respect to simultaneous detection
of vertices and convergences as well as identifying overall salience
in a field of disks with varying sizes.

Future developments could consider the respective convergence time
implied by each case, as well as the number of visited nodes and edges
(e.g.~\cite{Costa_know:2006}) along time.  Also interesting would be
the application of the proposed methodology for the analysis of images
of neuronal cells, allowing the identification of branches and
vertices while also enhancing the convergences of processes which is
typically found in those cells. The results for the tangent driven
random walks also suggest the development of a computationally
effective algorithm for saliency detection where \emph{all} image
pixels falling under the tangent lines defined by the edges would be
incremented.  It would also be interesting to consider other types of
random walks, including self-avoiding and self-attracting dynamics, as
well as the incorporation of additional driving constraints implied by
prior knowledge about the objects, memory, assigned tasks, etc.  Of
special importance would be the confrontation of the models with real
data provided by eye-tracking systems, which would allow the
identification of the most likely model parameters.

\vspace{1cm}

Luciano da F. Costa thanks Gustavo Vrech for the simulation of some
specific configurations and is also grateful to CNPq (301303/06-1) and
FAPESP (05/00587-5) for financial sponsorship.

\bibliographystyle{apsrev}
\bibliography{selatt}
\end{document}